\documentclass{article}
\usepackage{epsfig}

\newcommand{\nub}    {\overline{\nu}}
\newcommand{\shat}   {\mbox{$\mathrm{\hat{s}}$}} 
\newcommand{\roots}  {\mbox{$\sqrt{\mathrm s}$}}
\newcommand{\met}    {\mbox{$\mathrm{\not\!\!E_{T}}$}}

\newcommand{\zee}    {\mbox{$\mathrm Z \rightarrow\ \mathrm e^{+}\mathrm
e^{-}$}}

\newcommand{\zmumu}  {\mbox{$\mathrm Z \rightarrow\ \mathrm{\mu}^{+}\mathrm{\mu}^{-}$}}
\newcommand{\wenu}   {\mbox{$\mathrm W \rightarrow\ \mathrm e\mathrm \nu$}}
\newcommand{\wmunu}  {\mbox{$\mathrm W \rightarrow\ \mathrm{\mu}\mathrm{\nu}$}}
\newcommand{\sleq}   {\raisebox{-.6ex}{${\textstyle\stackrel{<}{\sim}}$}}
\newcommand{\sgeq}   {\raisebox{-.6ex}{${\textstyle\stackrel{>}{\sim}}$}}
\newcommand{\qqbar}  {\mbox{${q}\bar{q}$}} 
\newcommand{\ppbar}  {\mbox{${p}\bar{p}$}} 
\newcommand{\bbbar}  {\mbox{${b}\bar{b}$}}

\newcommand{\ptm}    {\mbox{$\mathrm{P_{T}}$}}
\newcommand{\etm}    {\mbox{$\mathrm{E_{T}}$}}

\newcommand{\mw}     {\mbox{$\mathrm {M_{W}}$}}

\newcommand{\ptl}    {\mbox{$\mathrm {P_{T}^{\mathit l}}$}}

\newcommand{\ptz}    {\mbox{$\mathrm{P_{T}^Z}$}}
\newcommand{\ptw}    {\mbox{$\mathrm{P_{T}^W}$}}

\newcommand{\mtr}    {\mbox{$\mathrm{M_{T}}$}}

\newcommand{\uvec}   {\mbox{$\mathrm{\vec{U}}$}}
\newcommand{\jpsi}   {\mbox{$\mathrm{J/\psi}$}}

\newcommand{\dzero}  {\hbox{DO\kern-0.62em\raise+0.2ex\hbox{/}}}
\newcommand{\pb}     {\mbox{${\rm pb}^{-1}$}}
\newcommand{\fb}     {\mbox{${\rm fb}^{-1}$}}

\def\Title#1{\begin{center} {\Large {\bf #1} } \end{center}}

\begin{document}

\Title{Electroweak Measurements from Hadron Machines}

\bigskip\bigskip

\begin{raggedright}  

{\it Mark Lancaster\index{Lancaster, M.}\\
Department of Physics and Astronomy\\ 
University College London\\
London, WC1E 6BT, UK.}
\bigskip\bigskip
\end{raggedright}

\section{Introduction}
The discovery of the W and Z gauge bosons~\cite{UA1_UA2} at the
S{\ppbar}S in 1983 marked the beginning of direct electroweak
measurements at a hadron machine. These measurements vindicated the tree
level predictions of the Standard Model. The new generation of hadron
collider machines now have data of such precision that the electroweak
measurements are probing the quantum corrections to the Standard
Model. The importance of these quantum corrections was recognised in the
award of the 1999 Nobel Prize~\cite{NOBEL_PRIZE}. These corrections are
being tested by a wide variety of measurements ranging from atomic
parity violation in caesium~\cite{CESIUM} to precision measurements at
the Z pole~\cite{LEP1_EWK} and above~\cite{LEP2_EWK} in $e^{+}e^{-}$
collisions. In this article, the latest experimental electroweak data
from hadron machines is reviewed. I have taken a broad definition of a
hadron machine to include the results from NuTeV~\cite{NUTEV} ($\nu$N
collisions) and HERA~\cite{H1,ZEUS} ($ep$ collisions) as well as the
results from the Tevatron ({\ppbar} collisions). This is not an
exhaustive survey of all results~\cite{REVIEW}, but a summary of the new
results of the past year and in particular those results which have an
influence on the indirect determination of the Higgs mass.  This article
will cover the direct determinations of the W boson and top quark masses
from {\ppbar} collisions at {\roots} = 1.8~TeV from the two Tevatron
experiments, CDF and {\dzero}. These results are based on $s$-channel
production of single W bosons and top quark pairs. The results presented
here from the NuTeV and HERA experiments allow one to make complementary
measurements and probe the electroweak interaction in the space-like
domain up to large momentum transfers in the $t$-channel.

\section{Data Samples}
 The first observation and measurements of the W boson were made at the
CERN S{\ppbar}S by the UA1 and UA2 experiments. These measurements were
based on modest event samples ($\sim$ 4~{\it k} events) and integrated
luminosity (12 \pb). Since that time the Tevatron and LEP2 experiments
have recorded over 1~\fb\ of W data. The Tevatron experiments have the
largest sample of W events : over 180,000 from a combined integrated
luminosity of $\sim$ 220~\pb. The LEP experiments, despite a very large
integrated luminosity ($\sim$ 15000~\pb\ total across all experiments),
have event samples substantially smaller than the Tevatron
experiments. The LEP2 W results~\cite{LEP2_WMASS_XSEC} presented at this
conference are based on event samples of $\sim$ 6~{\it k} events per
experiment. However, despite the smaller statistics of the W event
sample in comparison to the Tevatron experiments, the LEP2 experiments
ultimately achieve a comparable precision. On an event by event basis,
the LEP2 events have more information; in particular the LEP2
experiments can impose energy and momentum constraints because they have
a precise knowledge of the initial state through the beam energy
measurement. The NuTeV experiment at FNAL has a large
sample ($\sim 10^{6}$) of charged current events mediated by the
$t$-channel exchange of a W boson. This allows an indirect determination
of the W mass through a measurement of $\sin^2\theta_{\rm w}$. This is
done by comparing the neutral and charged current cross sections in
$\nu$Fe and $\overline{\nu}$Fe collisions. The event samples available
for electroweak tests at HERA are still rather modest and thus at
present their results do not attain the precision of the {\ppbar} and
{${\nu}N$} results. However, the ability to span a large range in
momentum transfers and have both $e^{+}p$ and $e^{-}p$ collisions
allows a number of unique electroweak observations to be made.  The
results from the Tevatron experiments on the top quark are now reaching
their conclusion. These measurements based on only $\sim$ 100 events
selected from over $10^{12}$ {\ppbar} collisions at the Tevatron
represent what it is possible to achieve with a robust trigger and
imaginative analysis techniques.

\section{W Boson and Top Quark Mass Measurements}
A precise W mass measurement allows a stringent test of the Standard
Model beyond tree level where radiative corrections lead to a dependence
of the W mass on both the top quark mass and the mass of the, as yet
unobserved, Higgs boson. The dependence of the radiative corrections on
the Higgs mass is only logarithmic whilst the dependence on the top mass
is quadratic. Simultaneous measurements of the W and top masses can thus
ultimately serve to further constrain the Higgs mass beyond the LEP1/SLD
limits and potentially indicate the existence of particles beyond the
Standard Model. Similarly, non Standard Model decays of the W would
change the width of the W boson. A precise measurement of the W width
can therefore be used to place constraints on physics beyond the
Standard Model.  The latest results on the W mass from the LEP2 and
Tevatron experiments are now of such a precision that the uncertainty on
the top mass is beginning to become the limiting factor in predicting
the mass of the Higgs boson.

\section{Latest Top Mass Measurements}
The top quark discovery at the Tevatron in 1995 was the culmination of a
search lasting almost twenty years. The top quark is the only quark with
a mass in the region of the electroweak gauge bosons and thus a detailed
analysis of its properties could possibly lead to information on the
mechanism of electroweak symmetry breaking. In particular, its mass is
strongly affected by radiative corrections involving the Higgs boson. As
such a measurement of the top mass with a precision $<$ 10~GeV can
provide information on the mass of the Higgs boson. The emphasis in top
quark studies over the past two years at the Tevatron has thus been to
make the most precise measurement of the top quark mass~\cite{TOP_MASS}.
Substantial progress has been made in bringing the mass uncertainty down
from $>$~10~GeV, at the point of discovery, to 5.1~GeV in 1999. In the
past year, CDF has revised it systematic error analysis in the
``all-hadronic'' event sample and both experiments have published an
analysis of the ``di-lepton'' event sample. The nomenclature of the
event samples refers to the decay chain of the top quarks.  At the
Tevatron, top quarks are produced in pairs predominantly by {\qqbar}
annihilation and each top quark decays $> 99.9 \%$ of the time to
$Wb$. If both Ws decay to {$qq^{\prime}$}, then the final state from the
top system is {$qq^{\prime}$}{$qq^{\prime}$}{\bbbar} and the event
sample is referred to as ``all-hadronic''. Conversely, the ``di-lepton''
event sample is realised by selecting a
{$l^+\overline{\nu}l^-\nu$}{\bbbar} final state, where both Ws have
decayed leptonically (to $e$ or $\mu$). The ``lepton+jet'' event sample
is one in which one W has decayed hadronically and one leptonically. The
precision with which the top mass can be measured with each sample
depends on the branching fractions, the level of background and how well
constrained the system is.  The all-hadronic sample has the largest
cross section but has a large background from QCD six jet
events (S/N $\sim 0.3$) whilst the di-lepton sample has a small
background (S/N $\sim$ 4) but suffers from a small cross
section and the events are ``under-constrained'' since they contain two
neutrinos. The optimum channel in terms of event sample size, background
level and kinematic information content is the lepton+jet
channel. Indeed this channel has a weight of $\sim$ 80~\% in the
combined Tevatron average. In the new ``di-lepton'' analysis, it is not
possible to perform a simple constrained fit for the top mass because
the solution, owing to the two neutrinos, is under-constrained. One thus
makes a comparison of the dynamics of the decay with Monte-Carlo
expectations e.g. the {\met} distribution and assigns an event weight to
each possible solution of the fit : two for each top decay corresponding
to the two-fold ambiguity in neutrino rapidity.

\begin{figure}[h]
\begin{minipage}{0.49\linewidth}
\psfig{file=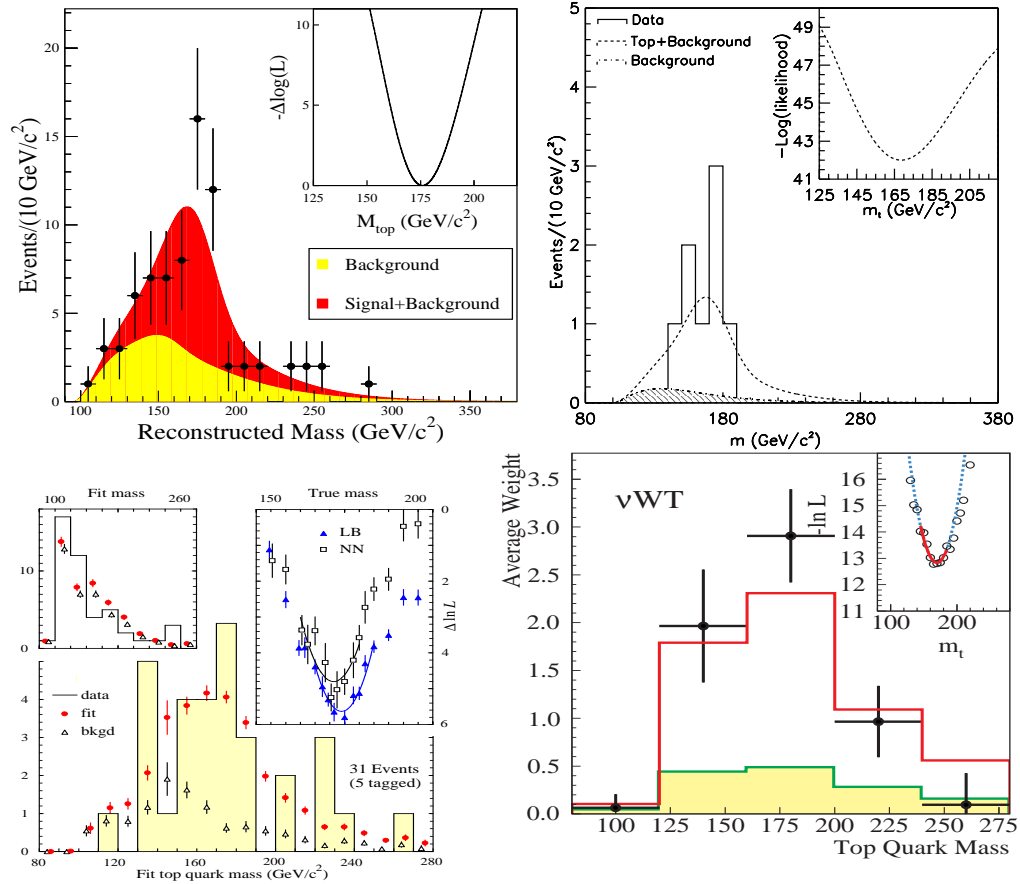,bb=15 240 560 780, 
       width=5.8in,height=5.0in,clip=}
\end{minipage}
\caption {The reconstructed top quark mass distributions in the
lepton+jet (leftmost plots) and di-lepton (rightmost plots) event
samples from CDF (upper plots) and {\dzero} (lower plots). The plots
also show the level of background and the likelihood fit distributions.
\label{FIGURE:FIG1}}
\end{figure}

The mass distributions of figure~\ref{FIGURE:FIG1} along with that from
CDF's analysis of the all-hadronic event sample have recently been
combined to provide a Tevatron average~\cite{TOP_AVERAGE} in which all
correlations between the various measurements have been carefully
accounted for. The two experiments have assumed a 100~\% correlation on
all systematic uncertainties related to the Monte Carlo models. Indeed,
the uncertainty in the Monte Carlo model of the QCD radiation is one of
the largest systematic uncertainties. This error source will require a
greater understanding if the top mass precision is to be significantly
improved in the next Tevatron run. The other dominant systematic error
is the determination of the jet energy scale which relies on using
in-situ control samples e.g. Z+jet, $\gamma$+jet events. In the next
run, due to significant improvements in the trigger system, both
experiments should be able to accumulate a reasonable sample of Z
$\rightarrow$ {\bbbar} events which will be of great assistance in
reducing the uncertainty in the $b$-jet energy scale.

\begin{figure}[h]
\begin{minipage}{0.49\linewidth}
\psfig{file=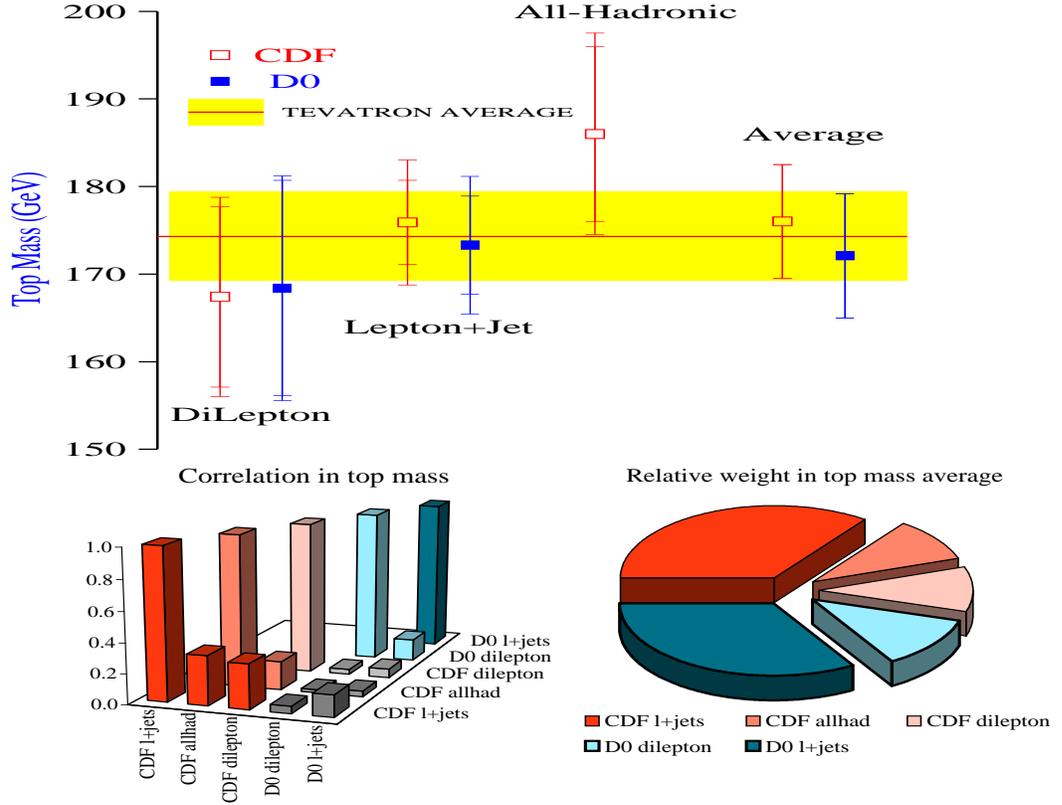,bb=65 220 600 850, 
       width=5.8in,height=5.0in,clip=}
\end{minipage}
\caption {UPPER : The top quark mass measurements of CDF and
{\dzero}. LOWER : The correlations and weighted contributions in the
final Tevatron average of the various event samples.
\label{FIGURE:FIG2}}
\end{figure}

The combined Tevatron mass value is 174.3 $\pm$ 3.2 (stat.) $\pm$ 4.0
(syst.) GeV. The individual mass measurements, the correlations between
them and the relative weights of the measurements in the average
are shown in figure~\ref{FIGURE:FIG2}.

This mass measurement is a supreme vindication of the Standard Model,
which, based on other electroweak measurements, predicts a top mass of :
172 $^{+14}_{-11}$~GeV. Of all the quark mass measurements the top quark
is now the best measured.
 
\section{Tevatron W Mass Measurements}

At the Tevatron W bosons are predominantly produced singly by quark
anti-quark annihilation. The quarks involved are mostly valence quarks
because the Tevatron is a {\ppbar} machine and the $x$ values involved in
W production (0.01 \sleq~$x$ \sleq\ 0.1) are relatively high. The W
bosons are only detected in their decays to $e\nu$ (CDF and \dzero) and
$\mu\nu$ (CDF only) since the decay to $qq^{\prime}$ is swamped by the
QCD dijet background whose cross section is over an order of magnitude
higher in the mass range of interest. At the Tevatron one does not know
the event \shat\ and one cannot determine the longitudinal neutrino
momentum since a significant fraction of the products from the {\ppbar}
interaction are emitted at large rapidity where there is no
instrumentation.  Consequently, one must determine the W mass from
transverse quantities~\cite{MT_THEORY} namely : the transverse mass
({\mtr}), the charged lepton \ptm\ ({\ptl}) or the missing transverse
energy ({\met}).  {\met} is inferred from a measurement of {\ptl} and
the remaining \ptm\ in the detector, denoted by {\uvec} i.e.
\[
\begin{array}[c]{ll}
\vec{\met} = - ( \uvec + \vec{\ptl}) & 
{\mathrm {~~~~and~{\mtr}~is~defined~as}} \\ 
\mtr = \sqrt{2\ptl\met(1-\cos\phi)} &  {\mathrm {~~~~where~
 \phi\ is~the~angle~between~}} \vec{\met}~ {\mathrm {and}}~ \vec{\ptl} 
\end{array}
\]
{\uvec} receives contributions from two sources. Firstly, the so-called W
recoil i.e. the particles arising from initial state QCD radiation from the
{\qqbar} legs producing the hard-scatter and secondly contributions from the
spectator quarks ({\ppbar} remnants) and additional minimum bias events
which occur in the same crossing as the hard scatter. This second
contribution is generally referred to as the underlying-event contribution.
Experimentally these two contributions cannot be distinguished. Owing to the
contribution from the underlying-event, the missing transverse energy
resolution has a significant dependence on the instantaneous {\ppbar}
luminosity. {\mtr} is to first order independent of the transverse momentum
of the W ({\ptw}) whereas {\ptl} is linearly dependent on {\ptw}. For this
reason, and at the current luminosities where the effect of the {\met}
resolution is not too severe, the transverse mass is the preferred quantity
to determine the W mass. However, the W masses determined from the
{\ptl} and {\met} distributions provide important cross-checks on the
integrity of the {\mtr} result since the three measurements have different
systematic uncertainties. 

The systematics of the LEP2 and Tevatron measurements are very different
and thus provide welcome complementary determinations of the W mass.
The systematics at LEP2 are dominated by the uncertainty in the beam
energy (which is used as a constraint in the mass fits) and by the
modeling of the hadronic final state, particularly for the events where
both W bosons decay hadronically. At the Tevatron, the systematics are
dominated by the determination of the charged lepton energy scale and
the Monte-Carlo modeling of the W production, in particular its \ptm\
and rapidity distribution. At the Tevatron, one cannot use a beam energy
constrain to reduce the sensitivity of the W mass to the absolute energy
(E) and momentum (p) calibration of the detector. Any uncertainty in the
detector E, p scales thus enters directly as an uncertainty in the
Tevatron W mass. This means that the absolute energy and momentum
calibration of the detectors must be known to better than 0.01\%. By
contrast at LEP, an absolute calibration of 0.5 \% is sufficient.

The W mass at the Tevatron is determined through a precise simulation of the
transverse mass line-shape, which exhibits a Jacobian edge at $\mathrm
{M_T \sim M_W}$. The simulation of the line-shape relies on a detailed
understanding of the detector response and resolution to both the
charged lepton and the recoil particles. This in turn requires a precise
simulation of the W production and decay. The similarity in the
production mechanism and mass of the W and Z bosons is exploited in the
analysis to constrain many of the systematic uncertainties in the W mass
analysis. The lepton momentum and energy scales are determined by a
comparison of the measured Z mass from \zee\ and \zmumu\ decays with the
value measured at LEP. The simulation of the W \ptm\ and the detector
response to it are determined by a measurement of the Z \ptm\ which is
determined precisely from the decay leptons and by a comparison of the
leptonic (from the Z decay) and non-leptonic \etm\ quantities in Z
events. The reliance on the Z data means that many of the systematic
uncertainties in the W mass analyses are determined by the statistics of
the Z sample.

The W and Z events in these analyses are selected by demanding a single
isolated high \ptm\ charged lepton in conjunction with missing
transverse energy (W events) or a second high \ptm\ lepton (Z
events). Depending on the analyses, the \met\ cuts are either 25 or 30
GeV and the lepton \ptm\ cuts are similarly 25 or 30 GeV. CDF only uses
\wenu\ and \wmunu\ events~\cite{CDF_WMASS} in the rapidity region :
$|{\eta}| < 1$, whereas \dzero~\cite{D0_WMASS} uses \wenu\ events out to
a rapidity of $\sim$ 2.5.  In total $\sim$ 84{\it k} events are used in the W
mass fits and $\sim$ 9{\it k} Z events are used for calibration.

\subsection{Lepton Scale Determination}

\begin{figure}[h]
\begin{minipage}{0.49\linewidth}
\psfig{file=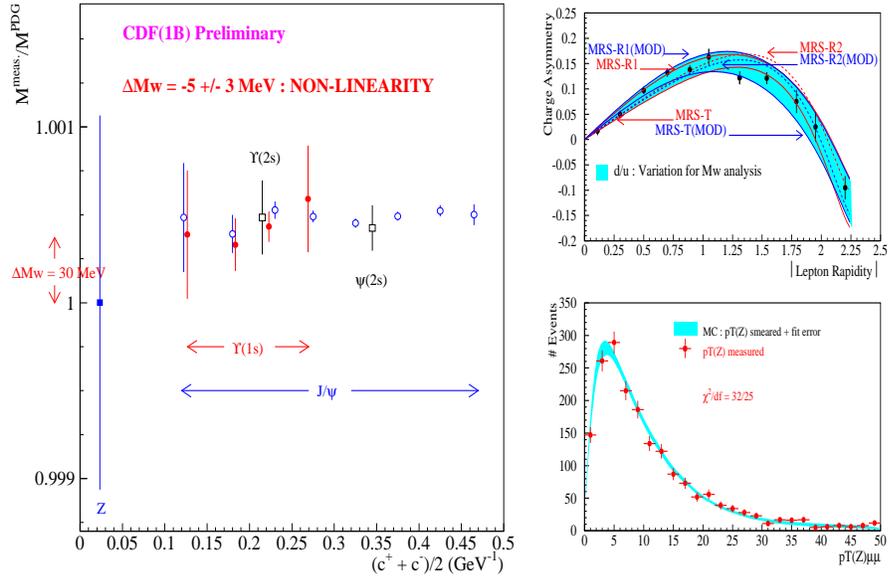,bb=20 190 565 485, 
       width=5.0in,height=3.2in,clip=}
\end{minipage}
\caption {LEFT: The CDF determination of the momentum scale and
         non-linearity using dimuon resonances.
         RIGHT UPPER: The modified PDF sets used in the Mw analysis,
         which span CDF's $W$ charge asymmetry measurement.
         RIGHT LOWER : The Z \ptm\ distribution as measured by CDF in
         the \zmumu\ channel. 
         \label{FIGURE:FIG3}}
\end{figure}

The lepton scales for the analyses are determined by comparing the
measured Z masses with the LEP values. The mean lepton \ptm\ in Z events
(\ptm\ $\sim$ 42 GeV) is $\sim$ 5~GeV higher than in W events,
consequently in addition to setting the scale one also needs to
determine the non-linearity in the scale determination i.e. to determine
whether the scale has any \ptm\ dependence. {\dzero} does this by
comparing the Z mass measured with high \ptm\ electrons with \jpsi\ and
$\pi^0$ masses measured using low \ptm\ electrons as well as by
measuring the Z mass in bins of lepton {\ptm}. In the determination of
CDF's momentum scale the non-linearity is constrained using the very
large sample of \jpsi\ $\rightarrow \mu\mu$ and $\Upsilon$ $\rightarrow
\mu\mu$ events which span the \ptm\ region : 2 $<$ \ptm\ $<$ 10~GeV. The
non-linearity in the CDF transverse momentum scale is consistent with
zero (see Fig.~\ref{FIGURE:FIG3}).  This fact in turn can be exploited
to determine the non-linearity in the electron transverse energy scale
through a comparison of the measured E/p with a MC simulation of E/p
where no \etm\ non-linearity is included.  The lepton scale
uncertainties form the largest contribution to the W mass systematic
error. The non-linearity contribution to the scale uncertainty is
typically $\sim$ 10\% or less.

The Z lineshape is also used by both experiments to determine the charged
lepton resolution functions i.e. the non-stochastic contribution to the
calorimeter resolution and the curvature tracking resolution in the case of
the CDF muon analysis.


\subsection{W Production Model} 

The lepton \ptm\ and \met\ distributions are boosted by the non zero \ptw\
and the \met\ vector is determined in part from the W-recoil products. As
such a detailed simulation of the \ptw\ spectrum and the detector response
and resolution functions is a necessary ingredient in the W mass analysis.
The W \ptm\ distribution is determined by a measurement of the Z \ptm\
distribution (measured from the decay leptons) and a theoretical prediction
of the W to Z \ptm\ ratio. This ratio is known with a small uncertainty and
thus the determination of the W \ptm\ is dominated by the uncertainty
arising from the limited size of the Z data sample. The \ptz\
distribution of the CDF \zmumu\ sample is shown in Fig.~\ref{FIGURE:FIG3}.
The detector response
and resolution functions to the W-recoil and underlying event products are
determined by both experiments using Z and minimum bias events. Since the
W-recoil products are typically produced along the direction of the vector
boson \ptm\ and the underlying event products are produced uniformly in
azimuth, the response and resolution functions are determined separately in
two projections -- one in the plane of the vector boson and one
perpendicular to it. Typically one finds the resolution in the plane of the
vector boson is poorer owing to the presence of jets (initial state QCD
radiation from the quark legs) which are absent in the
perpendicular plane where the resolution function matches closely that
expected from pure minimum bias events. The parton distribution functions
(PDFs) determine the rapidity distribution of the W and hence of the charged
lepton. Both experiments impose cuts on the rapidity of the charged lepton
and so a reliable simulation of this cut is necessary if the W mass
determination is not to be biassed. On average the u quark is found to carry
more momentum than the d quark resulting in a charge asymmetry of the
produced W i.e. W$^{+(-)}$ are produced preferentially along the ({\ppbar})
direction. Since the V-A structure of the W decay is well understood, a
measurement of the charged lepton asymmetry therefore serves as a reliable
means to constrain the PDFs. To determine the uncertainty in the W mass
arising from PDFs, MRS PDFs were modified to span the CDF charged lepton
asymmetry measurements~\cite{CDF_WASYM}. This is illustrated in
Fig.~\ref{FIGURE:FIG3}.

\subsection{Mass Fits}
\begin{figure}[h]
\begin{minipage}{0.49\linewidth}
\psfig{file=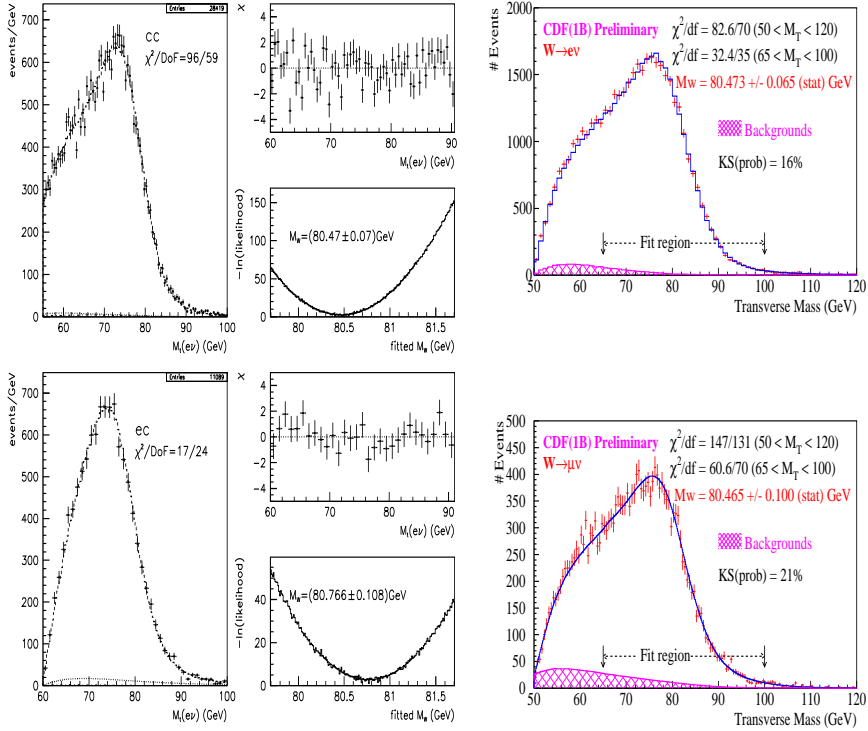,bb=40 100 530 430, 
       width=4.9in,height=4.2in,clip=}
\end{minipage}
\caption {Transverse mass distributions compared to the best fit. LEFT :
          {\dzero}'s published central-electron analysis and preliminary
	 end-cap analysis. RIGHT : CDF's electron and muon channel
         analyses. The fit likelihood  and residuals are also shown for
         the two {\dzero} distributions.
         \label{FIGURE:FIG4}}
\end{figure}

The W mass is obtained from a maximum likelihood fit of {\mtr} templates
generated at discrete values of \mw\ with $\Gamma_W$ fixed at the Standard Model
value.  The templates also include the background distributions, which
are small ($<$~5\%) and have three components : W $\rightarrow \tau\nu$,
followed by $\tau \rightarrow \mu/e\nu\nu$, QCD processes where one
mis-measured jet mimics the \met\ signature and the other jet satisfies
the charged lepton identification criteria and finally Z events where
one of the lepton legs is not detected. The transverse mass fits for the
{\dzero} end-cap electrons and the two CDF measurements are shown in
Fig.~\ref{FIGURE:FIG4}. 
The uncertainties associated with the measurements are listed in
Table~\ref{TABLE:ERRORS}.  The uncertainties of the published \dzero\
central-electron analysis are also listed.
\begin{table}[htbp]
\vskip 0.4cm
\begin{center}
\begin{tabular}{|l|c|c|c|c|} \hline

{Error Source} & {\dzero\ (EC)} & {\dzero\ (C)} & {CDF (e)} & {CDF ($\mu$)} \\ \hline
Statistical             & 70      &  105   &  65     & 100 \\ 
Lepton Scale+Resolution & 70      &  185   &  80     & 90  \\
\ptw\ + \met\ Model     & 35      &  50    &  40     & 40   \\
Other experimental      & 40      &  60    &  5      & 30   \\
Theory (PDFs, QED)      & 30      &  40    & 25      & 20   \\ \hline
Total Error             & 120     & 235    & 113     & 143  \\ 
Mass Value              & 80.440  & 80.766 & 80.473  & 80.465       \\ \hline
Combined Mass Values    &       
\multicolumn{2}{c|} {80.497 $\pm$ 0.098 GeV}   &
\multicolumn{2}{c|} {80.470 $\pm$ 0.089 GeV}          \\ \hline
\end{tabular}
\end{center}
\caption{The mass values and uncertainties of the CDF and {\dzero} W
mass analyses using the 1994--1995 data. The uncertainties are quoted in
MeV. The mass values when the
1992--1993 data are included become : 80.474 $\pm$ 0.093 GeV for \dzero\
and 80.430 $\pm$ 0.079~GeV for CDF. (EC) denotes the large rapidity
end-cap analysis and (C) denotes the central rapidity analysis.
\label{TABLE:ERRORS}}
\end{table}
For both experiments the
largest errors are statistical in nature, both from the statistics of
the W sample and also the statistics of the Z samples which are used to
define many of the systematic uncertainties e.g. the uncertainties in
the lepton energy/momentum scales and the W \ptm\ model.  The CDF and
{\dzero} measurements are combined with a 25~MeV common uncertainty
which accounts for the uncertainties in PDFs and QED radiative
corrections which by virtue of being constrained from the same source
are highly correlated. Together the two experiments yield a W mass value
of 80.450~GeV with an uncertainty of 63~MeV. For the first time, both
Tevatron experiments have measurements with uncertainties below 100~MeV
and the combined uncertainty is comparable with the LEP2 results
presented at this conference.

\section{W Mass Result from NuTeV}
Neutrino scattering experiments have contributed to our understanding of
electroweak physics for more than three decades. Early determinations of
$\sin^{2}\theta_{W}$ served as the critical ingredient to the Standard
Model's successful prediction of the W and Z boson masses. More precise
investigations in the late 1980's set the first useful limits on the top
quark mass. The recent NUTEV measurement of the electroweak mixing angle
from neutrino-nucleon scattering represents the most precise
determination to date.  The result is a factor of two more precise than
the previous most accurate $\nu$N measurement~\cite{CCFR}.  In deep
inelastic neutrino-nucleon scattering, the weak mixing angle can be
extracted from the ratio of neutral current (NC) to charged current (CC)
total cross sections. Previous measurements relied on the
Llewellyn-Smith formula, which relates these ratios to
$\sin^{2}\theta_{W}$ for neutrino scattering on isoscalar
targets~\cite{LLEWELLYN-SMITH}. However such measurements were plagued
by large uncertainties in the charm contribution (principally due to the
imprecise knowledge of the charm quark mass). An alternate method for
determining $\sin^2\theta_W$ that is much less dependent on the details
of charm production and other sources of model uncertainty is derived
from the Paschos-Wolfenstein quantity, $R^{-}$~\cite{PW}~:

\begin{equation}
R^{-} \equiv \frac{\sigma(\nu_{\mu}N\rightarrow\nu_{\mu}X)-
                   \sigma(\nub_{\mu}N\rightarrow\nub_{\mu}X)}
                  {\sigma(\nu_{\mu}N\rightarrow\mu^-X)-  
                   \sigma(\nub_{\mu}N\rightarrow\mu^+X)}  
= \frac{R^{\nu}-rR^{\nub}}{1-r}=\frac{1}{2}-\sin^2\theta_W
\end{equation}

\begin{figure}[h]
\begin{minipage}{0.49\linewidth}
\psfig{file=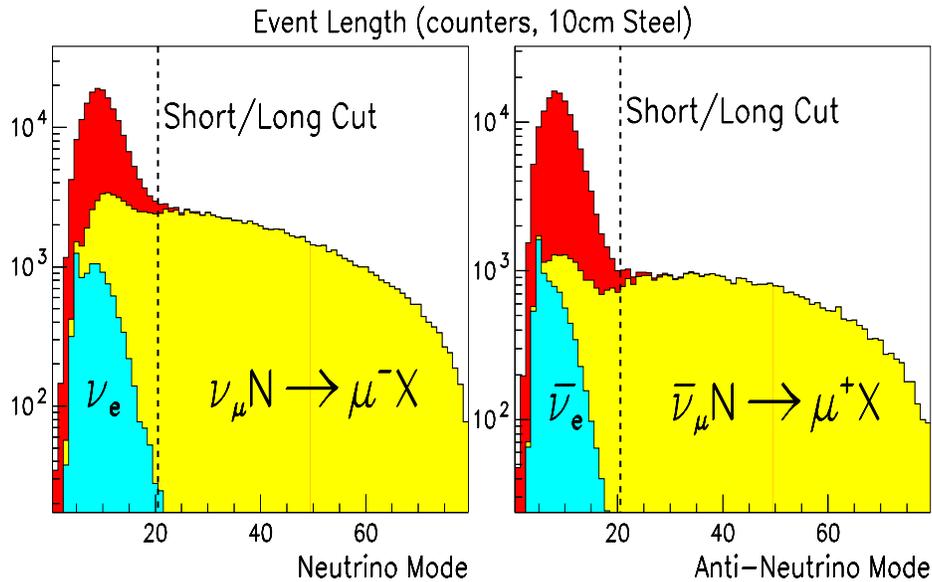,bb=0 0 580 290, 
       width=5.0in,height=3.2in,clip=}
\end{minipage}
\caption {The event length distributions (dark : NC events, light : CC
events) as a function of the number of counters for the $\nu$ and
$\overline\nu$ beams. The level of $\nu_e$ contamination is also shown.
\label{FIGURE:FIG5}}
\end{figure}

\noindent
Because $R^-$ is formed from the difference of neutrino and
anti-neutrino cross sections, almost all sensitivity to the effects of
sea quark scattering cancels. This reduces the error associated with
heavy quark production by roughly a factor of eight relative to the
previous analysis. The substantially reduced uncertainties, however,
come at a price. The ratio $R^-$ is difficult to measure experimentally
because neutral-current neutrino and anti-neutrino events have identical
observed final states. The two samples can only be separated via \emph{a
priori} knowledge of the incoming neutrino beam type. This is done by
using the FNAL Sign Selected Quadrupole Train (SSQT) which selects
mesons of the appropriate sign. The measured $\nub_{\mu}$ contamination
in the $\nu_{\mu}$ beam is less than 1/1000 and the $\nu_{\mu}$
contamination in the $\nub_{\mu}$ beam is less than 1/500. In addition,
the beam is almost purely muon-neutrino with a small contamination of
electron neutrinos ($1.3\%$ in neutrino mode and $1.1\%$ in
anti-neutrino mode). The NC and CC events are selected by their
characteristic event length : the CC events produce a muon and thus
register activity in the detector over a long length, whereas NC events
just produce a short hadronic shower. This is illustrated in
figure~\ref{FIGURE:FIG5} where the two event-length contributions to the
event sample are shown. The events are separated by a cut at the
20$^{\rm th}$ counter i.e. after $\sim$ 2~m of steel.

From the $\nu{N}$ interactions, 386~$k$ NC and 919~$k$ CC events are recorded
and from the $\overline{\nu}{N}$ sample 89~$k$ NC and 210~$k$ CC events. The
extracted value of $\sin^2\theta_W$(on-shell) = 0.2254 $\pm$ 0.0021
which can be translated into an Mw value of 80.26 $\pm$ 0.1 (stat.)
$\pm$ 0.05 (syst.)~GeV, where the systematic error also receives a
contribution from the unknown Higgs mass.

\section{Electroweak Results from HERA}
\begin{figure}[h]
\begin{minipage}{0.49\linewidth}
\psfig{file=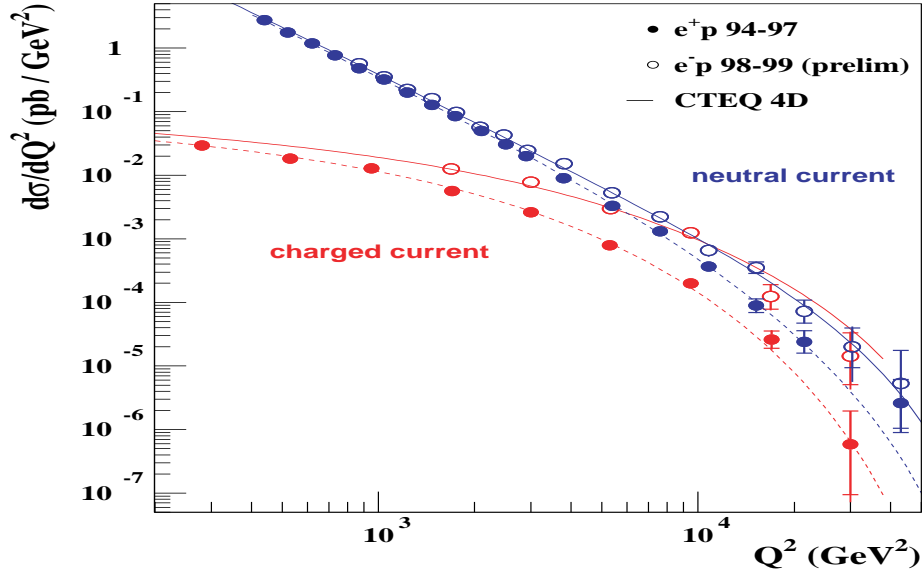,bb=29 164 541 668, 
       width=5.0in,height=3.2in,clip=}
\end{minipage}
\caption {The NC and CC $e^+p$ and $e^-p$ cross sections from ZEUS as a
function of $Q^2$.
\label{FIGURE:FIG6}}
\end{figure}

The two HERA experiments, ZEUS and H1, are now beginning to probe the
electroweak interaction in the space-like domain at scales of $\sim$
10$^{-3}$~fm. The results up to $Q^2$ values of 40,000 GeV$^2$ are in
good agreement with the Standard Model predictions. The use of both
$e^+p$ and $e^-p$ collisions and the fact that the experiments can
measure both neutral current and charged current processes, allows one
to directly observe electroweak unification in a single experiment. This
is illustrated in figure~\ref{FIGURE:FIG6} where both the neutral
current and charged current cross sections are shown as a function of
the Q$^2$. At low Q$^2$, the neutral current cross section, mediated by
$\gamma$ exchange, dominates; but as Q$^2$ increases one observes the
emergence of the charged current cross section (mediated by $W$
exchange) with a magnitude comparable to that of the neutral current
cross section (which becomes dominated by $Z_0$ exchange at high Q$^2$).
The detailed comparison of the four cross sections can be used to
measure parton distributions : in particular one can separate the light
quark contributions to the cross sections and determine the $u$ and $d$
quark distributions~\cite{MAX_KLEIN}. By using the helicity variable
$y$, where $y$ is related to the scattering angle in the electron-quark
center of mass frame, via : $(1-y) = \cos^2\frac{\theta}{2}$ and
measuring the NC cross section as a function of $(1-y)^2$ one observes
direct evidence for the $\gamma-Z_0$ interference contribution to the NC
cross section at high $Q^2$.

\begin{figure}[h]
\begin{minipage}{0.49\linewidth}
\psfig{file=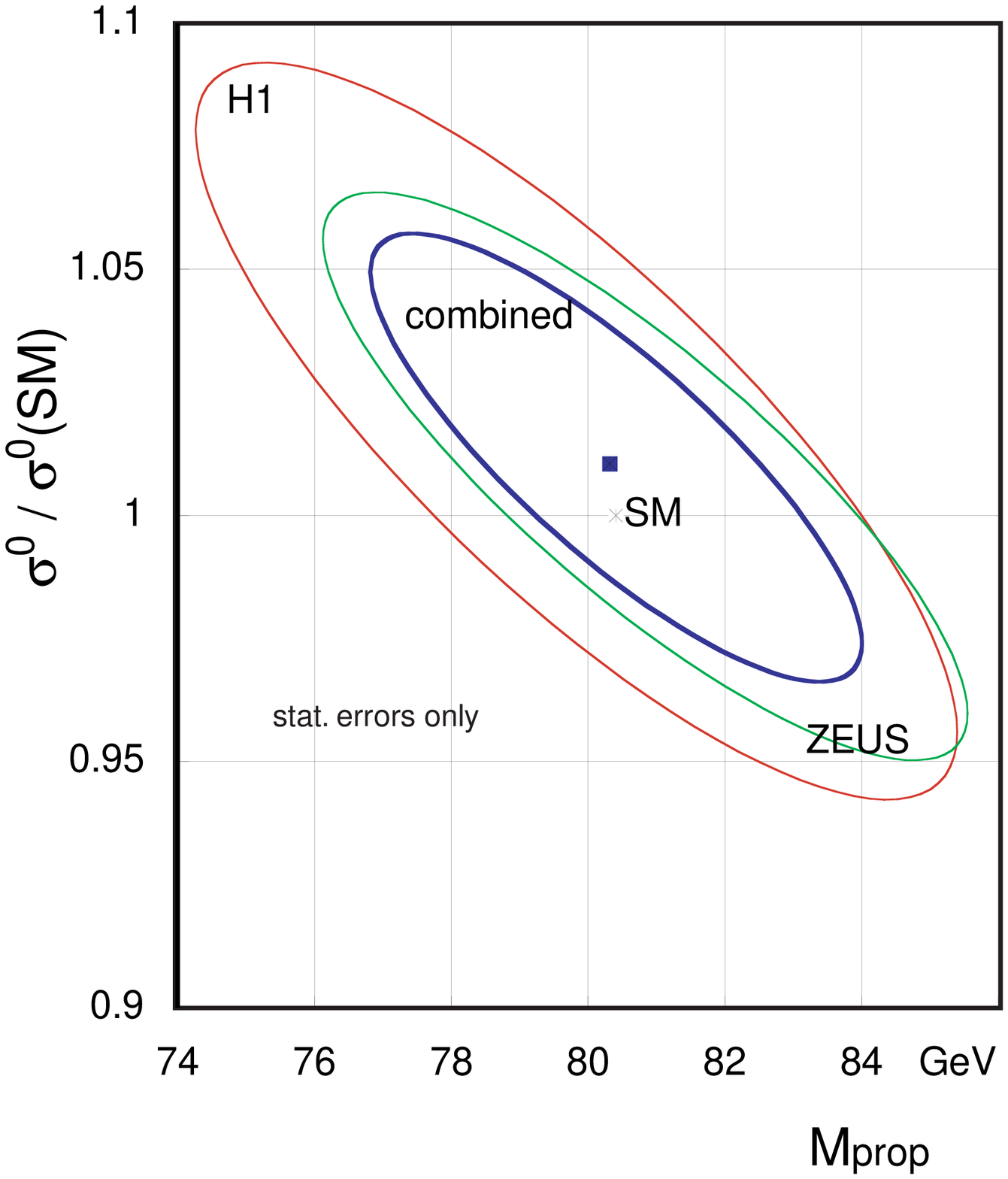,bb=22 110 520 690, 
       width=5.0in,height=3.2in,clip=}
\end{minipage}
\caption {The HERA charged current cross section compared to the SM
prediction as a function of the propagator mass, Mw.
\label{FIGURE:FIG7}}
\end{figure}

Although the HERA experiments have only produced a handful of direct Ws,
they have several thousand charged current events in which a virtual W
is exchanged in the $t$-channel. From a measurement of the Q$^2$
dependence of the cross-section one becomes sensitive to the charged
current propagator and hence W mass. This determination (see
figure~\ref{FIGURE:FIG7}) of the W mass agrees with the direct
determinations but is presently not competitive owing to the statistical
uncertainty and the systematics associated with parton distribution
functions : an uncertainty which can also be reduced with more
data. This determination is made more powerful if one also considers the
magnitude of the charged current cross section and assumes the Standard
Model relation between $G_F$ and Mw and Mz. This relation also receives
radiative corrections which depend on the masses of the fundamental
gauge bosons and fermions. One thus has a dependence on Mw in both the
$Q^2$ variation (propagator) and the cross section magnitude (via
$G_F$). By fixing the measured cross section to the Standard Model
value, one can obtain a W mass value with an uncertainty of $\sim$
400~MeV.

\section{Comparison of Mw results}

\begin{figure}[h]
\begin{minipage}{0.49\linewidth}
\psfig{file=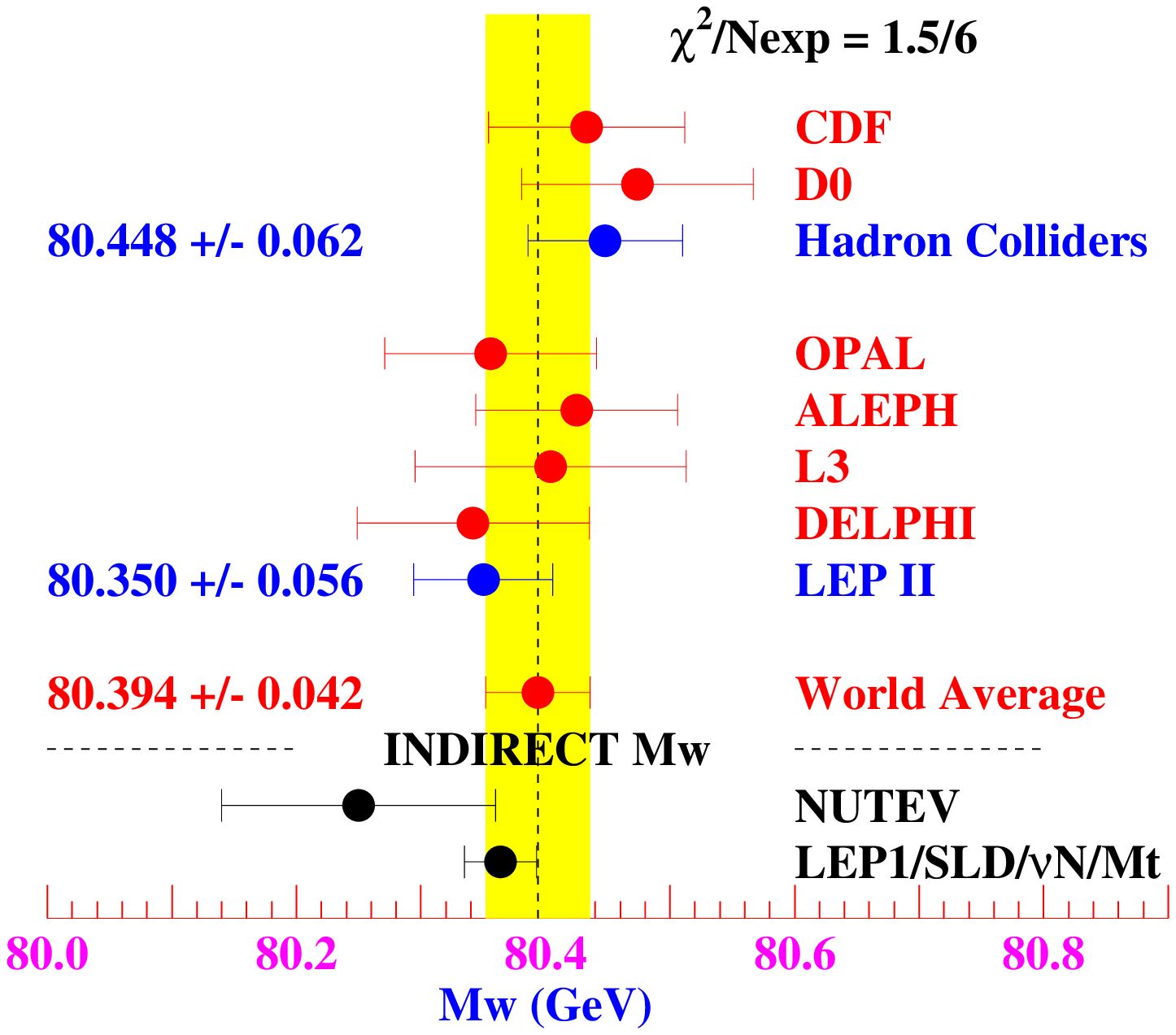,bb=15 15 530 430, 
       width=5.0in,height=3.2in,clip=}
\end{minipage}
\caption {The direct determinations of the W mass from the Tevatron and
LEP2 experiments are compared with the indirect measurement from NuTeV
and the prediction based on fits to existing electroweak data.
\label{FIGURE:FIG8}}
\end{figure}

The LEP2 mass values are compared with the Tevatron values in
figure~\ref{FIGURE:FIG8}. They are in excellent agreement despite being
measured in very different ways with widely differing sources of
systematic error. These direct measurements are also in very good
agreement with the indirect measurement from NuTeV and the prediction
based on fits to existing, non W, electroweak data.

\begin{figure}[h]
\begin{minipage}{0.49\linewidth}
\psfig{file=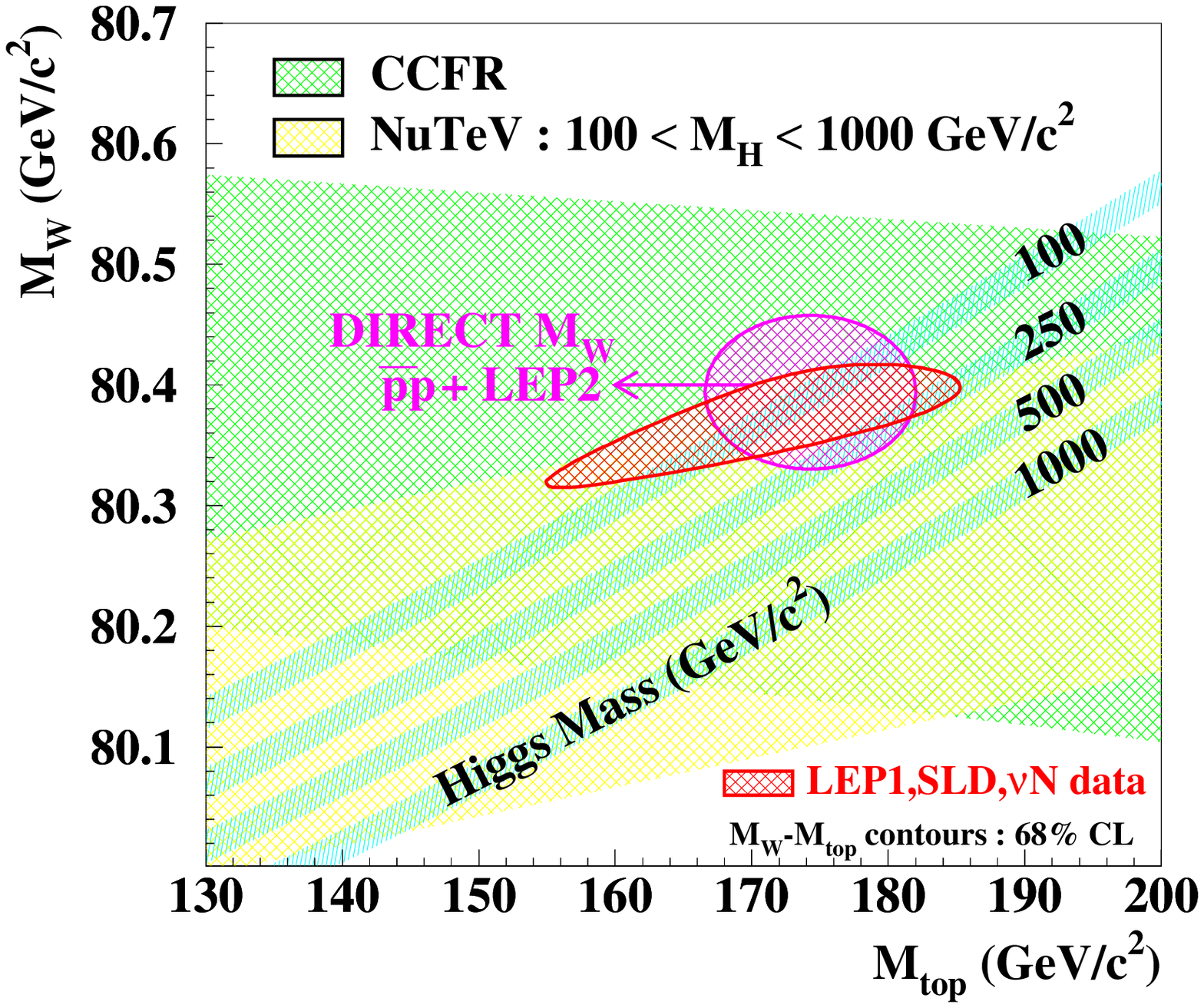,bb=30 180 560 650, 
       width=5.0in,height=3.2in,clip=}
\end{minipage}
\caption {The direct measurements of the W boson and top quark mass from
the Tevatron experiments are compared to the W mass measurements from
LEP2 and the predictions based on electroweak fits to LEP1/SLD/$\nu$N
data. The Standard Model predictions for the Higgs mass as a function of the W and
top masses are also shown.
\label{FIGURE:FIG9}}
\end{figure}

The precision of these measurements has increased the sensitivity that
one now has to the mass of the Higgs Boson. Indeed, it is now the
uncertainty on the top quark mass that is now becoming the limiting
factor in the determination of the Higgs mass.  As
figure~\ref{FIGURE:FIG9} shows, the available data tends to favour a
Higgs boson with a mass $<$ 250~GeV.

\section{W Width and branching fraction measurements}

\begin{figure}[h]
\begin{minipage}{0.49\linewidth}
\psfig{file=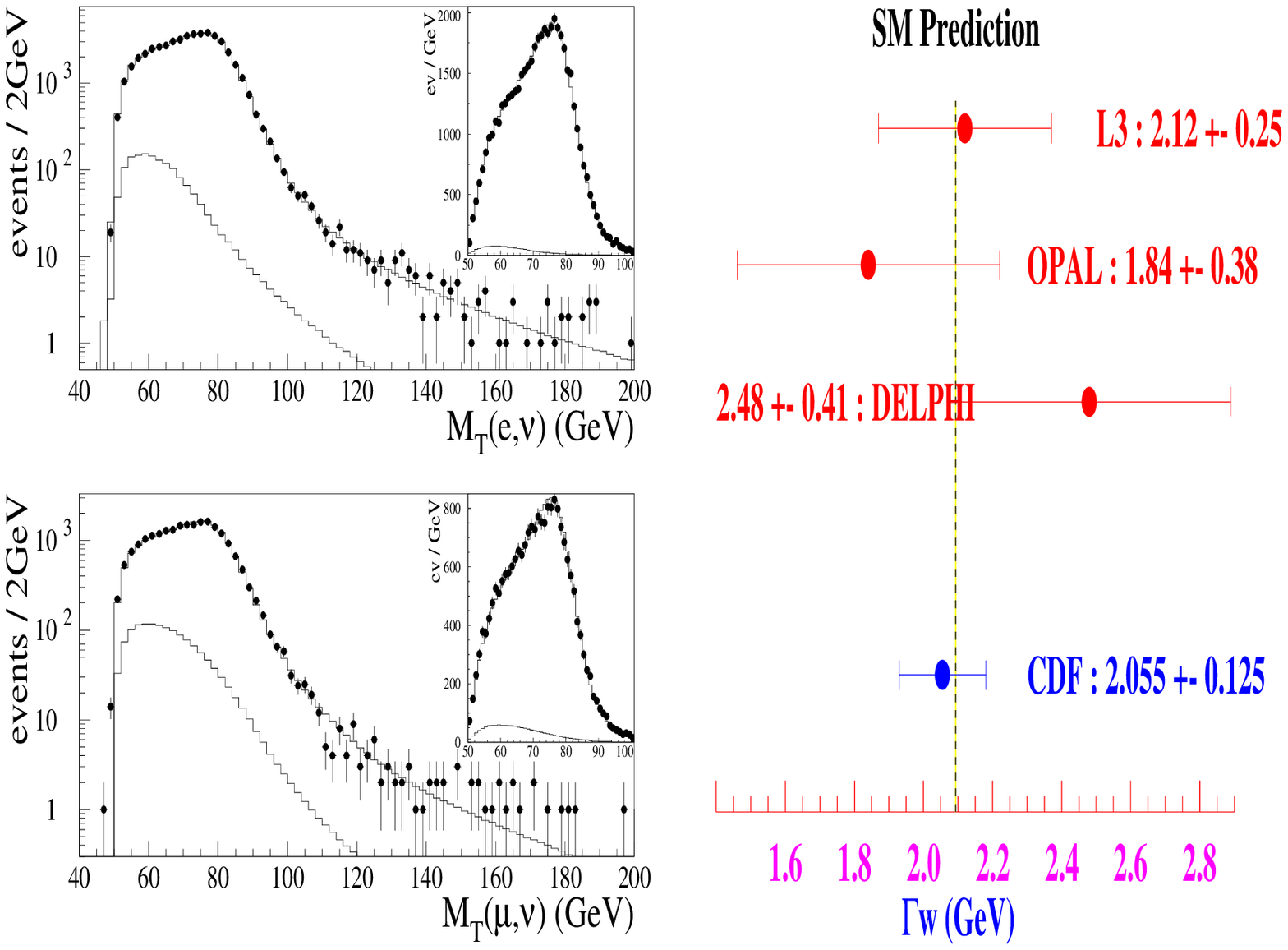,bb=30 220 570 640, 
       width=5.0in,height=3.2in,clip=}
\end{minipage}
\caption {LEFT: The transverse mass distribution of the CDF \wenu\
(upper) and \wmunu\ (lower) data showing the events at high transverse
mass from which the W width is determined. RIGHT: A comparison of the
direct W width determinations from LEP and the Tevatron with the SM
prediction.
\label{FIGURE:FIG10}}
\end{figure}

The Tevatron presently has the most precise direct determination of the
W width and has measurements of the W branching fractions comparable in
precision to the LEP2 measurements.  The Tevatron experiments determine
the width by a one parameter likelihood fit to the high transverse mass
end of the transverse mass distribution. Detector resolution effects
fall off in a Gaussian manner such that at high transverse masses (\mtr\
\sgeq\ 120~GeV), the distribution is dominated by the Breit-Wigner
behaviour of the cross section (see figure~\ref{FIGURE:FIG10}). In the
fit region, CDF has 750 events, in the electron and muon channels
combined.

At LEP2, the W branching fractions are determined by an explicit cross
section measurement whilst at the Tevatron they are determined from a
measurement of a cross section ratio. Specifically, the W branching
fraction can be written as :
$\sigma.{\rm BR}(W \rightarrow e\nu) =
\frac{\sigma_W}{\sigma_Z} \cdot \frac{\Gamma(Z \rightarrow
ee)}{\Gamma(Z)} \cdot \frac{1}{R};$ 
where 
${ R} = {{\sigma_W \cdot { Br(W \rightarrow e\nu)}} \over
        {\sigma_Z \cdot Br(Z \rightarrow ee)}} $ 
is the measurement made at the Tevatron. This determination thus relies
on the LEP1 measurement of the Z branching fractions and the theoretical
calculation of the ratio of the total Z and W cross sections. The
Tevatron measurement, $\sigma.{\rm BR}(W \rightarrow e\nu)$ = 10.43
$\pm$ 0.25~\% is now becoming systematics limited. In particular, the
uncertainty due to QED radiative corrections in the acceptance
calculation and in $\frac{\sigma_W}{\sigma_Z}$ contributes 0.19~\% GeV
to the total systematic uncertainty of 0.23~\%. The corresponding
measurement from LEP2 is $\sigma.{\rm BR}(W \rightarrow e\nu)$ = 10.52
$\pm$ 0.26~\%

The large sample of W events at the Tevatron has also allowed a precise
determination of $g_\tau/g_e$ through a measurement of the ratio of W
$\rightarrow \tau\nu$ to W $\rightarrow e\nu$ cross sections.  The
latest Tevatron measurement of this quantity is $g_\tau/g_e$ = 0.99
$\pm$ 0.024, in excellent agreement with the Standard Model prediction
of unity and the LEP2 measurement of $g_\tau/g_e$ = 1.01 $\pm$ 0.022.

\section{Outlook}

The majority of electroweak results presented here are presently
dominated by statistical uncertainties : either directly or in the
control/calibration samples. For NuTeV no further data is planned and
thus the precision of their measurement of $\sin^2\theta_W$ will remain
dominated by the statistical uncertainty. In contrast, both the Tevatron
and HERA are undergoing luminosity upgrades augmented with substantial
improvements in the collider's detectors.  At HERA, 150{\pb} per year
per experiment is expected, as is the availability of polarised
electrons and positrons in $ep$ collider mode.  At the Tevatron, at least
2{\fb} is expected per experiment at an increased center of mass energy
of 2~TeV. It is expected that both the top quark mass and W mass
measurements will become limited by systematic uncertainties.  The
statistical part of the Tevatron W mass error in the next run will be
$\sim$ 10~MeV, where this also includes the part of the systematic error
which is statistical in nature e.g. the determination of the charged
lepton E and p scales from Z events. At present the errors
non-statistical in nature contribute 25 MeV out of the total Tevatron
W mass error of 60 MeV. A combined W mass which is better than the final
LEP2 uncertainty can thus be anticipated~\cite{DURHAM_PROC}. The W width
is expected to be determined with an uncertainty of 20--40~MeV. The
statistical uncertainty, in the next Tevatron run, on the top quark mass
will be $\sim$~1~GeV. The systematic error arising from uncertainties in
the jet energy scale and modeling of QCD radiation are expected to be
the dominant errors, with a total error value of 2--3~GeV per experiment
expected. 

\section{Conclusions}

The hadron collider experiments continue to make significant
contributions to electroweak physics which complement those from
$e^+e^-$ machines. The higher cross sections and {\roots} allow some
unique measurements to be made e.g the mass of the top quark. The
precision of many measurements is comparable to, if not better than,
that achieved at $e^+e^-$ machines. All results : top mass, W mass, W
width, branching fractions, $\sin^2\theta_W$, cross sections etc are in
excellent agreement with the Standard Model in a range of
processes~:~$qq, eq, \nu{q}$ over a wide span in $Q^2$. The future is
bright and significant new results are expected from the Tevatron and
HERA experiments before the LHC.

\end{document}